\documentstyle[12pt,fullpage,epsfig]{article}
\begin{document}
\title{\bf Phase Relaxation of Electrons in Disordered Conductors}
\author{B.L. Altshuler$^{1,2}$, M.E. Gershenson$^3$, and
I.L. Aleiner$^4$
\ \\
\ \\
$^{1}$NEC Research Institute, 4
Independence Way, Princeton, NJ 08540\\
$^{2}$Physics Department,
Princeton University, Princeton, NJ 08544\\
$^{3}$Serin Physics
Laboratory, Rutgers University, Piscataway, NJ 08855\\
$^{4}$Physics and Astronomy Department,
SUNY at Stony Brook\\
Stony Brook, NY 11794}

\date{}
\maketitle
\ \\
{\bf Abstract}\\
\ \\
Conduction electrons in disordered metals and heavily
doped semiconductors at low temperatures preserve their phase
coherence for a long time: phase relaxation time $\tau_\varphi$ can
be orders of magnitude longer than the momentum relaxation time.
The large difference in these time scales gives rise to well
known effects of weak localization, such as anomalous
magnetoresistance. Among other interesting characteristics, study
of these effects provide quantitative information on the
dephasing rate $1/\tau_\varphi$. This parameter is of fundamental
interest: the relation between $\hbar/\tau_\varphi$ and the
temperature $T$ (a typical energy scale of an electron)
determines how well a single electron state is defined. We will
discuss the basic physical meaning of $1/\tau_\varphi$ in different
situations and its difference from the energy relaxation rate.
At low temperatures, the phase relaxation rate is governed by
collisions between electrons.  We will review existing theories
of dephasing by these collisions or (which is the same) by
electric noise inside the sample.  We also discuss recent
experiments on the magnetoresistance of $1D$ systems: some of
them show saturation of $1/\tau_\varphi$ at low temperatures, the
other do not.  To resolve this contradiction we discuss dephasing
by an external microwave field and by nonequilibrium electric
noise.

\section{Introduction}

\label{sec:1}

Anomalous magnetoresistance in disordered conductors (doped
semiconductors and metals) has been recognized for almost 50
years~\cite{MRhistory}.  For a long time this phenomenon has
remained a puzzle.  The theoretical understanding of the
anomalous magnetoresistance emerged as a spinoff of the theory of
Anderson localization. It turned out that the correction to the
conductivity, which is due to quantum interference at large
length scales, is very sensitive to weak magnetic fields. The
quantum correction itself may be much smaller than the classical
conductivity. Nevertheless, the weak field magnetoresistance is
dominated by this correction and its basic features (its
amplitude, dependence on both magnitude and direction of the
magnetic field, etc.) are very different from that of the
classical magnetoresistance.  Since the quantum correction can
eventually drive the system to the Anderson insulator, the regime
when this correction is small, is called the weak localization
(WL) regime, and the theory of the anomalous magnetoresistance is
now a part of the theory of weak localization.

A qualitative physical interpretation of WL is usually based on
estimation of the probability $P(t)$ to find a particle at given
time $t$ at the same point where it was at time $t=0$. The
quantum correction to this return back probability are due to the
interference between two amplitudes to return along the same
classical path in the opposite directions.  The quantum
correction is sensitive to the magnetic field that through the
loop formed by the classical self-returning path \cite{ALee}.
High sensitivity is associated with the fact that the typical
area $S$ of this loop is large.  For large loops, the area $S$ is
proportional to the length of the path, i.e., to the time $t$ it
takes for an electron to circle the loop.  Indeed, the motion is
diffusive provided $t$ exceeds the elastic mean free time $\tau$.
This means that the typical size of the loop is of the order of
$\left(Dt\right)^{1/2}$ and $S(t) \simeq Dt$, where
$D=v_F^2\tau/d$ is the diffusion constant ($d$ is the number of
dimensions, $v_F$ is the Fermi velocity of electrons).

Let us consider DC resistance of an ``infinite'' system. In this
case, there are two length scales which determine the relevant
size of the return path.  The first one is associated with the
magnetic field, while the second is determined by inelastic
collisions of electrons.

The effect of the magnetic field on the conductance is a
manifestation of the Aharonov -- Bohm effect~\cite{AB}.  Let us
consider the amplitude $ A(t)=\sum_j A_j(t)$ of the return
probability $P(t) = |A(t)|^{2}$, index $j$ here labels the
classical path.  In the absence of the magnetic field, the two
paths $j$ and $\bar{j}$ which differ only by direction are
characterized by the same classical action. Hence, in the
semiclassical approximation their contribution to $A(t)$ have
identical phases. This leads to constructive interference, to
an enhancement of $P(t)$ as compared with its classical value
$(Dt)^{d/2}$.  As a result, this interference reduces diffusion
constant $D$ and increases resistance $R$.

The equality of these two phases is a direct consequence of the
$T$ - invariance of the system, and is therefore violated by the
magnetic field.  The difference between the two phases is
determined by the magnetic flux encompassed by the returning
trajectories $\Phi_{(j)} (t) = HS_j(t)$
\begin{equation}
\phi_{j} - \phi_{\bar{j}} = 2\pi\frac{\Phi_{(j)}} {\Phi_{0}};\quad
\Phi_{0} = hc/2e,
\end{equation}
where $S_j$ is the directed area swept by the $j$th trajectory.
Thus, the interference contribution associated with $j$th path
acquires the oscillatory factor $\cos ( 2\pi\Phi_{(j)}/\Phi_0 )$.
Since $S_j$ are random, all the contributions from trajectories
sweeping typical area larger
than $
\simeq \Phi_{0}/H$ are diminished.  As the result, the first
characteristic length scale is the magnetic length $L_{H} =
\sqrt{\hbar c/eH}$.

Magnetoresistance will be determined by the classical
trajectories sweeping the area of the order of $L_{H}^2$,
provided electrons preserve the phase coherence during the
corresponding time $L_{H}^{2}/D$. Elastic collisions do not effect
this phase coherence, while inelastic interactions of the
electron with environment (other electrons, phonons, etc.) tend
to destroy it.  One can introduce a length scale $L_{\varphi}$ which
corresponds to substantial suppression of the coherence. This
scale and corresponding time $\tau_{\varphi} = L_{\varphi}^{2}/D$
depends on temperature $T$, and plays an important role in the
description of weak localization phenomena.  Since the
magnetoresistance depends on the ratio of $L_{\varphi}$ and $L_{H}$,
and $L_{H}$ can be tuned by changing the magnetic field,
magnetoresistance can be used for the study of inelastic
collisions and dephasing of electrons at low temperatures.

A conductor can be considered as infinite for our purposes as
long as the distance between the leads is much larger than at
least $L_{min}$ which is the smallest of the two characteristic
scales $L_{\varphi}$ and $L_{H}$.  It does not mean however, that
all of the dimensions of the system should be as large. The
sample has a dimension $d = 1,2,3$ depending on the relation
between its transverse dimensions and $L_{min}$.  We can speak
about $1d$ wires ($2d$ films) when both thickness of the wire $a$
and its width $W$ (thickness of the film $a$) is smaller than
$L_{min}$, despite the fact that both $a$ and $W$ are far in
excess of the electronic wavelength $k_{F}^{-1}$.  Since in this
lecture we discuss recent experiments on magnetoresistance of
$1d$ metallic wires and semiconductor structures, we will
consider here only the $1d$ case, though some of the statements
are of a more general validity.

The paper is organized as follows.  In Section~\ref{sec:2}, we
discuss qualitatively the main theoretical ideas on the dephasing
in disordered systems.  Readers interested in rigorous
derivations are urged to consult Ref.~\cite{AA}.
Section~\ref{sec:3} reviews recent observations of the saturation
of $\tau_\varphi$ at low temperatures Ref.~\cite{Mohanty}.  Authors of
Ref.~\cite{Mohanty} have suggested an explanation based on the
dephasing by the zero point motion of electrons.  We think that
this explanation is erroneous and elaborate on this issue more
in Sec.~\ref {sec:4}. Section~\ref{sec:5} describes new experiments
in which no saturation was observed, though the values of
$\tau_\varphi$ were much longer than the universal ``cut-off''
suggested.  As an explanation of the saturation in
Ref.~\cite{Mohanty}, proposed in Sec.~\ref{sec:6} the
dephasing by the external microwave radiation, and we show in
Sec.~\ref {sec:7} that the microwave radiation can efficiently
dephase electron without significant heating.

\section{Inelastic e-e Collisions and Dephasing Rate}

\label{sec:2}

In early papers (see {\em e.g.} Ref.~\cite{Thouless}) on the
theory of localization, the dephasing rate $1/\tau_{\varphi}$ was
considered to be of the same order as the inelastic collision
rate in perfectly clean conductors.  The latter can be expressed
as the sum of the electron - phonon $1/\tau_{e-ph}\simeq
T^{3}/\Theta_{D}^{2}$ and electron - electron $1/\tau_{e-e}\simeq
T^{2}/E_{F} $ contributions, where $E_{F}$ and $\Theta_{D}$ are
the Fermi and Debye energies correspondingly (here and almost
everywhere below we put $ \hbar =1$ and $k_{B} = 1$).  Even under
this assumption, the $e-e$ contribution dominates at low enough
temperatures.  It became clear later that static disorder
strongly enhances the $e-e$ contribution to the inelastic scattering
rate \cite{Schmid,AA}, while $1/\tau_{e-ph}$ is less affected
\cite{Reyzer}.  As the result, both dephasing and energy
relaxation rates at low temperatures are governed by collisions
between electrons.

To recall the main results on the $e-e$ dephasing rate, let us
start with a single electron excitation, assuming that $T = 0$,
and the rest of the electron gas occupies states below the Fermi
level.  Dependence of $1/\tau_{e-e}$ on the energy of the
excitation (energy of an electron counted from the Fermi level)
$\epsilon$ can be determined in a perturbative calculation~\cite
{Schmid,AA,inelastic}.  The result (Eq.(4.4) of
Ref.~\cite{AA}) can be rewritten through the dimensionless
conductance $g(L)$ (conductance measured in units of $e^{2}/h
\simeq 1/25.8k\Omega $) of a d - dimensional cube of the size $L$
\begin{equation}
\frac{1}{\tau_{e-e}(\epsilon)} = C_{d} \frac{\epsilon}{g(L_{\epsilon})},
\quad L_{\epsilon} \equiv \sqrt{D/\epsilon},  \label{eq:2.1}
\end{equation}
where $C_{d}$ is the dimension-and-coupling-constant-dependent
coefficient.  For a weakly interacting $1d$ electron gas $C_{d} =
\sqrt{2}$. Eq.(2) can also be rewritten as
\begin{equation}
\frac{1}{\tau_{e-e}(\epsilon)} = C_{d} \delta_{1}(L_{\epsilon}),
\label{eq:2.2}
\end{equation}
where $\delta_{1} (L) = (L^{d} \nu)^{-1}$ is a one-particle mean
level spacing in a $d$ - dimensional cube of the size $L \cdot
\delta_{1}$ is determined by one-particle density of states $\nu$.

There are several interpretations of this result.  One of them~
\cite{AA} is based on the concept of the interaction time which
becomes much longer in the disordered case due to diffusive
rather than ballistic motion of electrons. It is also possible to
appeal to statistical properties of exact one-electron wave
functions~\cite{Blanter,AGKL,AP}, and we outline this
interpretation below.

Inelastic rate $1/\tau_{e-e} (\epsilon)$ is determined by a pair
of collisions between electrons with all four energies - two
initial ($\epsilon > 0 $ and $\epsilon^\prime <0 $) and two final
($\epsilon - \omega > 0$ and $ \epsilon^\prime + \omega > 0 $) -
belonging to the energy strip with the width $2\epsilon$ centered
at the Fermi level, all the energies here are counted from the
Fermi level.

Given the typical absolute value $M(L, \omega, \epsilon,
\epsilon^\prime)$ of the matrix element for such a collision in a
sample with a size $L$, the inelastic rate can be
estimated~\cite{Aronov,AGKL} with the help of the Fermi
Golden Rule
\begin{equation}
\frac{1}{\tau_{e-e}} \propto \sum_{0 < \omega < \epsilon} \sum_{-\omega
<
\epsilon^\prime < 0} \frac {M(L, \omega, \epsilon, \epsilon^\prime
)^{2}}{\delta_{1} (L)},  \label{eq:2.3}
\end{equation}

The matrix elements can be represented as integrals of products of
four exact one particle wave functions. In a disordered system,
these wave functions oscillate randomly in space, and are only
weakly correlated with each other.  As a result, the matrix
elements are random and for $L$ smaller than $ L_{\epsilon}$ (0D
case), their typical absolute value $M(L)$ turns out to be of the
order of $\delta_{1}(L)/g(L)$, where the small factor $g^{-1}$
reflects the weakness of the correlation between the wave
functions. (In the limit $g \rightarrow \infty$ Random Matrix
Theory is valid; according to this theory, there is no correlation
at all between different eigenvectors and the non-diagonal matrix
elements vanish.) Each sum in Eq.~(\ref{eq:2.3}) leads to the
factor $\sim \epsilon /\delta_{1}(L)$. As a result, in $0D$ case
$ 1/\tau_{e-e}$ can be estimated as~\cite{AP}
\begin{equation}
\frac{1}{\tau_{e-e}} \simeq \frac{\epsilon^{2}}{g^{2} \delta_{1}(L)}
\label{eq:2.4}
\end{equation}

It increases with $\epsilon$, and at $L = L_{\epsilon}$, the rate
$1/{ \tau_{e-e}}$ becomes of the order of
$\delta_1(L_{\epsilon})$.  This estimate corresponds exactly to
Eqs.~(\ref{eq:2.1}) and (\ref{eq:2.2}), and it remains valid even
for large samples, $L > L_{\epsilon}$, since $ 1/\tau_{e-e} $
cannot depend on $L$ in this limit.

When making estimate (\ref{eq:2.4}) we assumed that
$1/\tau_{e-e}$ in Eq.~(\ref {eq:2.1}) is determined by the energy
transfer $\omega$ of the order of $ \epsilon$.  To take into
account quasielastic processes with small energy transfer, let us
find the dependence of the matrix element on the transmitted
energy $ \omega$. From comparison of Eqs.~(3) and (4) this 
dependence in $d$ -- dimensional sample reads
\begin{equation}
  M^{2} \sim \frac{\delta_{1} (L)^{3}\delta_{1}
    (L_\omega)}{\omega^2} = \frac{
    \delta_{1}(L)^{4}L^{d}}{\omega^2 L_{\omega }^{d}} \propto
  \omega^{-2+d/2}.
\label{eq:2.5}
\end{equation}
This energy dependence of the matrix element reflects the
properties of noninteracting disordered system and is not
sensitive to the distribution function.

We see from Eq.~(\ref{eq:2.5}) that the matrix elements diverge
when $\omega \rightarrow 0$ for $d < 4$. At $T=0$ this divergence
is not dangerous because of two summations in Eq.~(\ref{eq:2.4}).
However, the situation changes when the temperature is finite. In
this case $|\epsilon \prime|$ in Eq.~(\ref{eq:2.4} ) is
determined by ${\rm max}\{T,\omega\}$, and at $\omega < T$
summation over $\epsilon \prime$ can be substituted by the factor
$T/\delta_{1}(L)$.  According to Eq.~(\ref{eq:2.5}), it means
divergence of the sum over $\omega$ in the lower limit for
$d=1,2$.  Therefore, {$1/\tau_{e-e}$ is ill-defined at finite
  temperatures and in low dimensions}~\cite{AndersonLee}.

This is not a catastrophe, though: $1/\tau_{e-e}$ itself has no
physical meaning.  When the energy relaxation rate
$1/\tau_{\epsilon}$, ({\em i.e.} the inverse time of
thermalization of an excitation with energy $ \epsilon$ much
larger than temperature $T$) is considered, the quasielastic
processes are not important. Therefore Eqs.~(\ref{eq:2.1}) and
(\ref{eq:2.2} ) give a good estimate of $1/\tau_{\epsilon}$.
The phase relaxation rate $ 1/\tau_{\varphi}$ is more delicate and
requires additional consideration~\cite {AAK,AA}, since it
involves the electron with typical energy $T$, which quasielastic
scattering rate is divergent.

An additional phase caused by an inelastic collision is just a
product of the energy transfer $\omega$ and the time $t$ that
passed after the collision. It means that collisions with
arbitrary small $\omega$, which give negligible contribution to
$1/\tau_{\epsilon}$, can cause dephasing, provided the phase is
detected over a sufficiently long time.  It is clear that the
typical observation time $t$ just can not be larger than the
dephasing time itself $\tau_{\varphi}$. Therefore in cases when
$1/\tau_{(e-e)}$ diverges ($d = 1,2$), the divergence should be
cut off by $\omega \sim 1/\tau_{\varphi}$. As a result, instead of
Eqs.~(\ref{eq:2.1}) and (\ref{eq:2.2} ), we obtain a
self-consistent equation for $\tau_{\varphi}$ and $L_{\varphi}$:
\begin{equation}
\frac{1}{\tau_{\varphi}(T)} = C_{d} \frac{T}{g(L_{\varphi})} = C_{d}
\delta_{1}(L_{\varphi}) T\tau_{\varphi}, \quad L_{\varphi} \equiv
\sqrt{D\tau_{\varphi}}.
\label{eq:2.6}
\end{equation}

\noindent Solving Eq.~(\ref{eq:2.6}) we find \cite{AA,AAK}
\begin{equation}
  \frac{1}{\tau_{\varphi}}=\left(T^2\Delta_\xi\right)^{1/3}, \quad
  L_\varphi = \xi \left(\frac{\Delta_\xi}{T}\right)^{1/3}
  \label{Lphi}
\end{equation}
in $1D$ case.  Here energy $\Delta_\xi= D/\xi^2=
2\pi^2\delta_1(L)/g(L)$, has the meaning of the level spacing on
the localization length $\xi$ (we assume $g(L) \gg 1$).
Corresponding result in two dimensions is
\[
\frac{1}{\tau_{\varphi}} = \frac{T}{g}\ln g,
\]
Therefore, $\tau_{\varphi} \rightarrow \infty$, when $T \rightarrow
0$.  It should be noted that Eq.~(\ref{eq:2.6}) for
$\tau_{\varphi}$, as well as the WL theory as a whole, is valid when and
only
when $g(L_{\varphi})$ is large.  The dephasing length $L_{\varphi}$
increases when $T \rightarrow 0$.  As soon as it reaches the
localization length $\xi$, Eq.~(\ref{eq:2.6}) can not be used any
more: $ g(\xi ) \sim 1$ by definition. On the other hand, as long
as $g(L_\varphi) \gg 1$, the
system behaves as a Fermi liquid, since  $T\tau_\varphi \gg 1$.

It is also important to emphasize that the condition of the validity
of the WL approach, $g(L_\varphi) \gg 1$ {\em does not impose any
restriction on the total conductance of the wire}, 
$g(L) = g(L_\varphi)(L/L_\varphi)$. {\em E.g.}, for the samples described in
Sec.~\ref{sec:5}, $g(L) \simeq 3 \times 10^{-3}$, whereas $g(L_\varphi) >
3$, so that WL consideration is still applicable.

We presented here a rather simplified interpretation of old
results, which were derived rigorously about 15 years ago and
were re-derived later in several ways~\cite{Stern,Chakravarty}.

\section{Experiments on Gold Wires}

\label{sec:3}

These old results have been recalled in connection with recent
measurement of the WL magnetoresistance of $1D$ $Au$
wires\cite{Mohanty}. From this measurements authors extracted the
temperature dependence of the dephasing rate $1/\tau_{\varphi}(T)$.
They have found that $\tau_{\varphi}(T)$, increases with cooling,
when $T$ is large enough, but at $T \sim 1K$ it saturates at a
level of about 1 nanosecond.

In fact, such saturation has been observed by many experimental
groups.  However, this apparent contradiction with the theory was
attributed to one of the two reasons. The saturation was
explained either by overheating the electrons (due to applied
voltage or to external noise) or by scattering of electrons by
localized spins. Authors of Ref.~\cite{Mohanty} have demonstrated
experimentally in a convincing and elegant way that both reasons
for the $\tau_{\varphi}(T)$ saturation are not applicable for their
samples.

First of all, they observed temperature dependence of the
resistivity -- probably, due to the effects of the interaction
between electrons -- at $T$ as low as $40 mK$. The very fact that
such dependence does exist is a strong evidence for the electron
gas to have the same temperature as the bath.  The effect of
paramagnetic impurities was ruled out by {\it adding} certain
concentration of $Fe$ into gold and observing how effect of these
additional localized spins disappears with cooling due to the
Kondo effect (inelastic cross-section on one-channel Kondo
impurities $\propto T^2$ at $T$ smaller than the Kondo
temperature).

These arguments convinced the authors of Ref.~\cite{Mohanty} that
the finite dephasing at $T=0$ is a fundamental and unavoidable
consequence of the interaction between electrons. In light of
this experimental data the saturation of $\tau_{\varphi}(T)$ at $T
\rightarrow 0$ appears to be a problem, so serious, that it
inspired several attempts to reconsider the foundations of the
theory of disordered conductors in the weak localization regime.

The main puzzle in experiments Ref.~\cite{Mohanty} is that
$\tau_{\varphi}(T)$ saturates when conductance is still very large:
$g(L_{\varphi}) \sim 10^{3}$.  It means that corresponding
zero-temperature dephasing length $L_{\varphi}(T=0)$ is much smaller
than the localization length $\xi$. Therefore, {\it assuming that
  this relation between $L_\varphi$ and $\xi$ always holds,
  one should conclude that any interaction between electrons
  rules out localization of quantum states in a weakly disordered
  wire with a finite cross-section.}

The authors of Ref.~\cite{Mohanty} came up with an analytical
estimate of $ L_{\varphi}(T=0) = L_{MJW}$. They argued that their
experimental data, as well as all other data available are
consistent with the estimate, $L_{\varphi}(T=0) \sim L_{MJW}$.  For a
wire with a thickness $a$ and a width $W$ their expression for
$L_{MJW}$ can be rewritten in the form
\begin{equation}
L_{MJW} = Wak_{F} = N/k_{F} = \xi/(k_{F}l),  
\label{bred}
\label{eq:3.1}
\end{equation}
where $k_{F}$ is Fermi wavenumber, $N = Wa/k_{F}^2$ is the
number of channels in the wire, and $l$ is the mean free path of
the electrons.  Note that in the WL regime $k_{F}l \gg
1$, and the localization length $ \xi$ can be estimated as $\xi
\sim lN$. Assuming that Eq.~(\ref{eq:3.1}) gives correct upper
limit for $L_{\varphi}$, one concludes that interactions between
electrons prevent localization, provided $N > 1$ and $k_{F}l >
1$.

In fact, one can discuss a possibility that for some reason
interaction between electron, in addition to inelastic dephasing,
causes static violation of the $T$ - invariance, {\em e.g.}
orbital ferromagnetism. This violation would saturate the temperature
dependence of the magnetoresistance. In this hypothetical case,
localization is possible and would be similar to the $1D$
localization of non-interacting electrons in the presence of
magnetic field (unitary ensemble). It would mean, though that the
localization length does not depend on the magnetic field at all
(in contrast to the experimental evidence [23], see below).

Therefore, assuming that the fact that $L_{\varphi}$ saturation is a
fundamental law, one ends up with the conclusion that 1D
localization either does not exist in WL regime, or it is
magnetic-field-independent.

\section{Nyquist -- Johnson Noise and Zero-Point Oscillations}

\label{sec:4}

Dephasing caused by the electron - electron quasielastic
collisions Eq.~(\ref{eq:2.6}), which we interpreted through
general properties of matrix elements, can be also understood
from a slightly different point of view~ \cite{AAK}. Instead of
thinking about many colliding electrons, we can consider one
electron, which while moving around the loop, is  subject (in
addition to the quenched disordered potential) to a random time
and space dependent electric field. This field is created by the
rest of the electron gas and is nothing but the {\it equilibrium
  electric noise} inside the conductor. The advantage of this
approach is that correlation functions of this field at large
times and distances are determined solely by the conductivity of
the system.

It turned out to be possible to take the dephasing effect of this
noise into account in a non-perturbative way, and to determine
the quantum correction to the conductivity $\sigma (H,T)$ as a
function of magnetic field and temperature. In $1D$ case
\begin{equation}
  \delta g\left(L > L_{\varphi}\right) \propto \frac{L_{\varphi}}{L}
  \frac{1}{ \left[\ln
Ai\left(\frac{(WL_{\varphi}H)^{2}r}{\varphi_{0}^{2}}\right)\right]^\prime
    }, \label{eq:3.2}
\end{equation}
where $[\ln Ai(x)]^\prime $ is the logarithmic derivative of the
Airy function, $L_{\varphi}$ is a determined by Eq.~(\ref{Lphi}) and
$r$ is geometry dependent coefficient of order unity.

Since Eq.~(\ref{Lphi}) is in qualitatively contradiction to the
experimental results, authors of Ref.~\cite{Mohanty} made an
attempt to explain dephasing at $T=0$ by zero-point oscillations
of the electric field~\cite{Mohanty2}, {\em i.e.} they have assumed
that dephasing is determined by the processes with the {\it
  energy transfer $\omega$ much larger than temperature.} Their
consideration resulted in the length scale $L_{MJW}$.

We do not believe that zero point oscillations can cause any
dephasing.  Indeed, one can naively consider any environment as a
set of harmonic oscillators.  This set is characterized by the
distribution of frequencies and couplings with a given quantum
particle~\cite{Caldeira}. A collision between the particle and an
oscillator is inelastic, provided the particle either transfers
energy to the oscillator and excites it, or receives energy from
the oscillator. The energy of the zero-point oscillations can not
be transferred, since this energy is simply a difference between
the ground state energy of the oscillator and the bottom of the
harmonic potential. At $T$ much smaller than the frequency of the
oscillator $\omega_{I}$, inelastic collisions are
impossible: the oscillator is in the ground state, and the
particle does not have enough energy to excite the oscillator.
Therefore, with probability exponentially close to unity, the
collision is elastic and the oscillator has as little chance to
cause dephasing as any static impurity.

We have to address the question again: why is the experimentally
observed dephasing length is {\it always} (as it is pointed out
in Ref.~\cite{Mohanty}) smaller than $L_{MJW}$? The answer is: it
is {\em not always the case}! In the next section we briefly discuss
experiments where dephasing lengths much larger than $ L_{MJW}$
have been observed.

\section{New Data on $L_{MJW}$ Decoherence in 1D $\delta $-doped GaAs
Wires.}

Recently, new data on the temperature dependence of $L_\varphi$
have been obtained for sub-micron-wide ``wires'' fabricated from
the $\delta $-doped GaAs structures \cite{gersh3}.  In these
samples, a single $\delta $-doped layer with concentration of Si
donors $N_D=5\times 10^{12}cm^{-2}$ is $0.1$ $ \mu m$ beneath the
surface of an undoped GaAs. The 1D wires were fabricated by
electron beam lithography and deep ion etching. A 50-nm-thin
silver film deposited on top of the structure was used as a
``gate'' electrode: the electron concentration $n$ and the
resistance of the samples can be ``tuned'' by varying the gate
voltage $V_g$ (for more details, see \cite {gersh3}). Below we
discuss the data obtained for the sample comprising 360 wires
connected in parallel; the length $L$ of each wire is 500 $\mu
m$, the effective wire width $W$ = 0.05 $\mu m$. Relatively high
concentration of carriers ensures that the number of occupied 1D
sub-bands is large ($\sim 10$ ) in the wires. The mean free path
of electrons $l$ increases with $n$ from $ 17$ $nm$ to $58$ $nm$;
($k_Fl\approx 6\mbox{---} 30$, where $k_F$ is the Fermi wave number).
The sample is {\it one-dimensional} with respect to the quantum
interference effects at low temperatures:$\ W<L_\varphi (T)\leq
\xi$. These 1D conductors demonstrate the Thouless
crossover~\cite{Thouless} from weak localization (WL) to strong
localization (SL) with decreasing the
temperature~\cite{gersh1,gersh2}; the crossover temperature $T_0$
can be varied over a broad range by the gate voltage.  In strong
magnetic fields, the crossover ``shifts'' toward lower
temperatures; this shift is accompanied with {\it doubling} of
the localization length, and {\it halving} of the hopping
activation energy in the SL regime~\cite{gersh1,gersh2}.

Both the temperature and magnetic field dependences of the
resistance of these samples are consistent with the theory of WL
and interaction effects on the ``metallic'' side of the crossover
($T>T_0$)~\cite{gersh3}. The phase coherence length $L_\varphi $
has been estimated from the WL magnetoresistance; the procedure
of extraction of $L_\varphi $ for 1D conductors has been
described in detail in \cite{gersh3,gersh0}. The dependences
$L_\varphi (T)$ are shown in Fig.~1 for different values of the
gate voltage $V_g$. The experimental values of $L_\varphi $ are
well described Eq.~(\ref{Lphi}).  over the whole temperature range that
corresponds to the WL regime.  The theoretical dependences
Eq.~(\ref{Lphi}) 
are extended in Fig.~1 down to the crossover temperature. The
dependence $L_\varphi (T)$ {\it do not saturate} down to the
crossover temperature and the quasiparticle description holds
over the {\it whole} WL temperature range.

The observed dependences $L_\varphi (T)$ {\it argues against }the
idea of the decoherence due to zero-point fluctuations of the
electric field~\cite {Mohanty}. Indeed, Eq.~(\ref{bred}) implies that {\it
  a}) narrow channels fabricated from 2DEG {\it cannot}
demonstrate 1D quantum corrections to the conductivity, since for
a strip of 2DEG, $L_{MJW}$ simply equals to the trip width W, and
{\it b}) the localization-induced crossover {\it should not} be
observable in such channels ($L_\varphi $, being limited by
$L_0=W$, is always much smaller than $\xi \gg W$ in this case).
The existence of the ``cut-off'' time $L^{2}_{MJW}/D$ would also
preclude observation of the {\it interaction-driven} 1D crossover
with decreasing the temperature. Indeed, as soon as $\tau
_\varphi $ approaches $\tau _0$, the broadening of the electron
energy levels, $\frac \hbar {\tau _\varphi }$, becomes
temperature-independent. It is worth noting that at $D
\hbar/L_{MJW}^{2} > k_BT$ the Fermi-liquid description should also
break down.

\begin{figure}[ht]
\epsfig{file=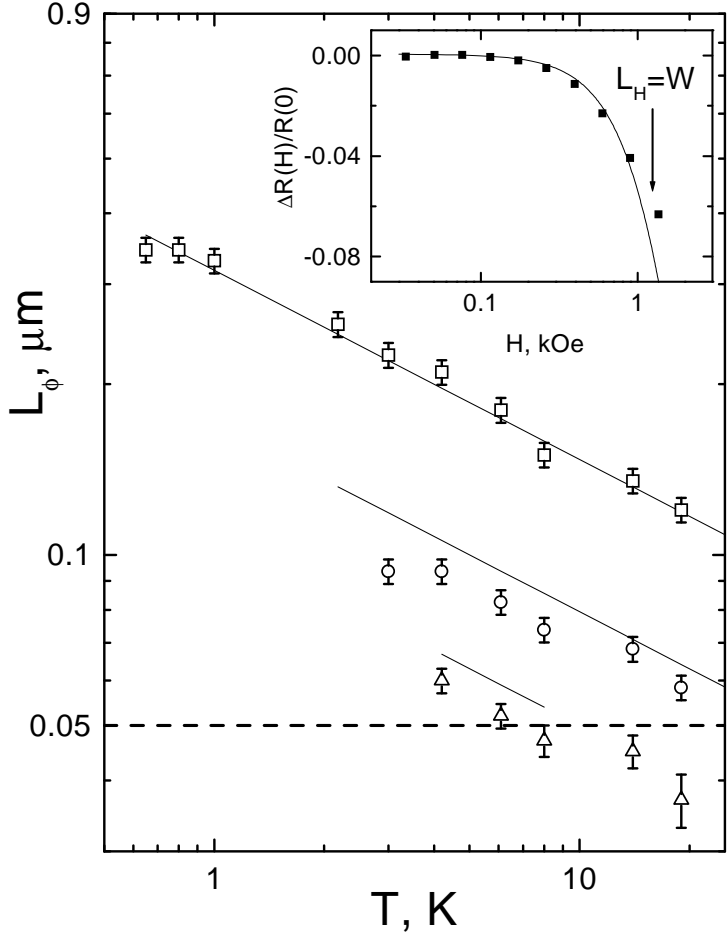, height=0.9\textwidth}
\caption{The phase coherence length versus temperature for a $\delta
  $-doped GaAs wire (for parameters, see text) at different
  $V_g$: $\Box $
$-0.7V$; $\
\bigcirc -0V$; $\Delta --0.35V.$ Solid lines - Eq.~(\protect\ref{Lphi}), the
dashed line - Eq.~(\protect\ref{bred}). 
The insert shows the magnetoresistance
at $T=8K$, $V_g=0V $, the solid line - the WL theory fit.}
\label{Fig.1}
\end{figure}

Both consequences of Eq.~(\ref{bred}) contradict available experimental
data.  First, the 1D corrections to the resistance of narrow
channels fabricated from 2DEG have been studied for more than a
decade. Second, the Thouless crossover has been observed in
$\delta $-doped wires with the ratio $\xi /W$ as large as
16\cite{gersh3}. For the data shown in Fig.~1, $L_\varphi $ near
the crossover exceeds the estimate $L_0$ by a factor of ~$\sim
$ 7 for $ V_g=0.7$ $V$; hence, the maximum experimental values of
$\tau _\varphi$ exceed $L^{2}_{MJW}/D$ by a factor of 50.  The
experimental values of $\tau _\varphi $ an order of magnitude
greater than the estimate Eq.~(\ref{bred}) have been also observed in Ref.
\cite{link1}. Thus, there is a strong experimental evidence that
saturation of the $L_\varphi (T)$ dependence is not intrinsic and
the mechanism of this saturation is not related to the
zero-temperature fluctuations of the electric field.\label{sec:5}

\section{Dephasing by High Frequency Radiation and External Noise}

\label{sec:6}

We believe that saturation of the dependence $L_\varphi $($T)$
observed in the experiments Ref.~\cite{Mohanty} is due to phase
breaking by the {\it external } microwave electromagnetic noise.
In order to explain our viewpoint, let us recall another old
story about weak localization. Suppose that we have applied to
our sample AC electric field (microwave radiation with some
frequency $\Omega $). The question we want to address is: how
will this radiation affect DC conductance and its dependence on
magnetic field?  Obviously, the radiation can heat the sample,
and the temperature dependence of the conductance will transform
into its dependence on the amplitude of the AC electric field
$E_{AC}$. However, it turns out that the dephasing effect of
radiation can be much more important than this
heating~\cite{AAKHFR}.

The dephasing effect of the microwave radiation depends on both
its amplitude $E_{AC}$ and frequency $\Omega$. According to
Ref.~\cite{AAKHFR} this mechanism of dephasing is not effective
when frequency is too low or too high. At $\Omega \rightarrow
\infty$ one can take the radiation into account using
perturbation theory, which gives $1/\tau_{\varphi}\propto \Omega
^{-2}$. The fact that very high-frequency radiation is not effective
in dephasing is easy to understand since such field is averaged out
in a course of diffusive motion of electron during time $\tau_\varphi$. 

In the opposite limit of low frequencies one can consider
dephasing by the electric field, which is linear in time $E(t)
\sim E_{AC}\Omega t$, since the phases, which correspond to two
directions of circling the loop remain equal even if a DC field
is present. Consider a closed trajectory with the return time $t$.
Let us divide this trajectory into small segments $j$; each
segment has the length of the order of the elastic mean free path
$l$.  Let the electron circle around the loop in, say, clockwise
direction pass segment $j$ at the time moment $t_j$; then, the
electron traveling along the same trajectory but in
counterclockwise direction will pass the same segment at a moment
$ t-t_j$.  Therefore, when passing the same segment in different
directions, the electron will acquire the different energies; the
difference in this energies $\delta\epsilon_j$ can be estimated
as
\begin{equation}
  \delta \epsilon_j = \alpha_j l e\left[E(t_j)-E(t-t_j)\right]
  \simeq \alpha_j eE_{AC}\, l\Omega\,(2t_j-t), \label{est1}
\end{equation}
Here $\alpha_j$ is a random number which depends on the angle
that the electric field makes with the direction of the electron
path over the region $j$ $ \langle \alpha_j \rangle = 0; \
\langle \alpha_i\alpha_j \rangle \simeq \delta_{ij} $. In order
to find the total energy difference acquired by the electron
during the time interval $\left[0;\, t_j\right]$, we have to add
up contributions from different segments $\epsilon_j =\sum_{0 <
  t_i < t_j} \delta \epsilon_i$ and in order to find the total
accumulated phase, we have to integrate the energy difference
over time
\begin{equation}
\delta \varphi (t) = \tau \sum_{0 < t_j < t}\epsilon_j= \tau \sum_{0 < t_j
< t}
\sum_{0 < t_i < t_j} \delta\epsilon_i.  \label{est3}
\end{equation}
Accumulated phase (\ref{est3}) is the random quantity, and,
therefore, it is represented by $\langle\delta\varphi^2\rangle$.
Substituting Eq.~(\ref{est1}) into Eq.~(\ref{est3}) and averaging
its square with the help of $\langle \alpha_i\alpha_j \rangle
\simeq \delta_{ij}$, we obtain
\begin{equation}
\langle \delta \varphi (t)^2\rangle \simeq \left(eE_{AC}\, l \Omega
t\right)^2
\tau^2 \left(\frac{t}{\tau}\right)^3 =D(\Omega eE_{AC})^{2}t^{5}.
\label{est4}
\end{equation}
Determining $\tau_{\varphi}$ from the equation $\langle\delta \varphi
(\tau_{\varphi})^2\rangle \sim 1,$ we find that at small frequencies $
1/\tau_{\varphi}$ increases with $\Omega$:
\begin{equation}
\frac{1}{\tau_\varphi} \simeq D^{1/5}\left(\Omega eE_{AC} \right)^{2/5}
\label{est5}
\end{equation}
All this consideration holds provided $\Omega\tau_\varphi \leq 1$.
At larger frequencies, as we have already mentioned, the
dephasing rate decreases very fast with frequency. Rate
$1/\tau_{\varphi}$ reaches its maximum at a certain frequency
$\Omega_{E}$, which is determined by $E_{AC}$
\begin{equation}
  \left[\frac{1}{\tau_{\varphi}(\Omega)}\right]_{max} = \frac{1}{
    \tau_{\varphi}(\Omega_{E})} \sim \Omega_{E}, \quad \Omega_{E}
  \approx (e^{2}E_{AC}^{2}D)^{1/3}.  \label{tauphimin}
\end{equation}

These theoretical predictions have been verified experimentally
on inversion channels at the $Si$ surface ~\cite{Kwon} and on
$Mg$ films ~\cite{Lindelof}. In this papers it was found that
conductance depends substantially on the power of microwave
radiation. The frequencies used were $\Omega = 9.1 {\rm MHz}$
~\cite{Kwon} and $\Omega = 0.66 {\rm GHz}, 3.61 {\rm
  GHz}$~\cite{Lindelof}.  This dependence had nothing to do with
heating (in Ref.~\cite{Kwon} the effect had even the opposite
sign than that heating would cause) and was in a reasonable
agreement with the theory ~\cite{AAKHFR}.

It is clear that any external noise should have dephasing effect
similar to the one of the microwave radiation. {\it We believe
  that it is external nonequilibrium noise that causes saturation
  of the dephasing rate in the experiments Ref.~\cite{Mohanty}},
while at $T> 1K$ the equilibrium (and, thus,
temperature-dependent) Nyquist -- Johnson noise determines $\tau
_\varphi $ . In the next section we compare dephasing and heating
effects of the external noise in the ${\rm GHz}$ frequency range
for a particular case of the $Au$ wires Ref.~\cite{Mohanty}.

\section{Dephasing and heating in $Au$ wires.}

\label{sec:heating} \label{sec:7}

Let us first estimate power of the microwave radiation sufficient
for phase breaking at the time scale $\tau _\varphi $ in a $1D$
conductor. We assume that the radiation is in the optimal
frequency range $\Omega \tau _\varphi \sim 1$.  As we have seen, it
also means that $\Omega _E\tau _\varphi \sim 1$ or, according to
Eq.~(\ref{tauphimin}), $e^2E_{AC}^2D\tau _\varphi ^3\sim 1$.  The
radiation dominates dephasing as soon as $1/\tau _\varphi $
given by this estimate exceeds the rate provided by the
equilibrium noise $\sim (T^2D/\xi ^2)^{1/3}$, see
Eq.~(\ref{Lphi}). For a wire with a length $L$ and a large
conductance $g(L)\gg 1$ this happens when $eE>T/[Lg(L)]$, since
the localization length can be written as $\xi \sim Lg(L)$. In
terms of the power $P_{AC}=\left( LE_{AC}\right) ^2/2R$, where
$R=h[e^2g(L)]^{-1}$ is the resistance of the wire, this
inequality takes the form
\begin{equation}
  P_{AC}>P_\varphi =R\left( \frac{ek_BT}\hbar \right)
  ^2=\frac{2\pi }{\hbar g(L)} (k_BT)^2.  \label{dephasingpower}
\end{equation}
Here we restore the Planck and Boltzmann constants. Note, that
this power is proportional to the total resistance of a wire.
Because of a very large resistance of the $\delta $-doped wires
studied in \cite {gersh1,gersh2,gersh3} ($R(4K)\simeq 9M\Omega $
for a single wire at $ V_g=0.7V$ in Fig.1), the microwave power
required for decoherence in such samples is rather large:
$P_\varphi \simeq 4\cdot 10^{-9}W$ at $T=1K$.  However, for 1D Au
wires with a small $R\simeq 0.3\mbox{---} 1.8k\Omega $~\cite{Mohanty},
$P_\varphi $ should be smaller by four orders of magnitude.

Does this microwave power heat the wire? The answer depends on
how efficiently the extra energy is removed from the sample. One
can propose two mechanisms of cooling: {\it a}) phonon emission
and {\it b}) heat flow along the sample into ``cold'' leads. Let
us start with the second mechanism, which is more important at
low temperatures even for rather long wires.  The expression for
the power removed from the wire due to the hot-electron
out-diffusion can be obtained using the Wiedemann-Franz
law~\cite{prob}.  Let $T$ be the temperature of the leads, and
$T_e$ the temperature of electrons in the wire.  (The
electron-electron interaction is sufficiently strong in thin
metal films at low temperatures to ensure thermalization of
electrons and to justify the approximation of local electron
temperature \cite{gersh4}). Calculations of the temperature rise
for different power levels and sample parameters can be found in
Ref.~\cite{mittal}: for a small absorbed power, the difference
$\Delta T=T_e-T$ has a parabolic profile along the wire, with the
peak equal to $1.5$ of $\Delta T$ averaged along the wire.
The estimate of the heat flow out of both ends of the wire into
the ''cold'' leads for small $\Delta T<<T$~has been done by
Prober\cite {prob}:
\begin{equation}
  P_{es}\simeq \left( \frac{2\pi k_BT}e\right) ^2\frac{T\Delta
    T}R={2\pi g(L)} \frac{k_B^2T\Delta T}\hbar \label{poweres}
\end{equation}

Comparison of Eq.~(\ref{dephasingpower}) with Eq.~(\ref{poweres})
shows that at overheating is very small at $P_{AC}\sim P_\varphi$
even in the absence of other cooling mechanisms:
\begin{equation}
\frac{\Delta T}T=g(L)^{-2}\ll 1.  \label{deltaT}
\end{equation}
Thus, for any sample with $g(L)>>1$, {\it the microwave radiation
  can efficiently destroy the phase coherence of the electron
  wavefunction without heating the electron gas. }

In fact, formula (\ref{deltaT}) even overestimates $\Delta T$ for
long wires.  In this case, the contribution of the phonon
emission becomes dominant.  For a sufficiently thin film, the
``bottleneck'' for the energy transfer from electrons to the
thermal bath at low temperatures is formed by the electron-phonon
interaction: the non-equilibrium phonons escape ballistically
into the substrate~\cite{gersh4}. In this case, the rate at which
energy flows out of the electron gas by phonon emission,
$P_{e-ph}$, is given by the expression:
\begin{equation}
P_{e-ph}=\frac{C_e}{\tau _{eph}\left( T\right) }\Delta T.  \label{eph}
\end{equation}
Here $C_e=(a\cdot W\cdot L)\gamma T_e$ is the heat capacity of
the electron gas in a wire of volume $a\cdot W\cdot L$, $\gamma $
is the Sommerfield parameter ($\simeq 70J/m^3K^2$ for Au), and
$\tau _{eph}$ is the inelastic electron-phonon scattering time.
Equation~(\ref{eph}) is valid when $\Delta T=T_e-T$, where $T$ is
the temperature of equilibrium phonons, is much smaller than both
$T_e$ and $T$. For an estimate of $\tau _{eph}$, we can use
recent results for the electron-phonon scattering time in thin
$Au$ films: $\tau _{eph}(T)\simeq 1\,{\rm ns}\times \,(1K/T)^2$
\cite{bel}.

\begin{figure}[ht]
\epsfig{file=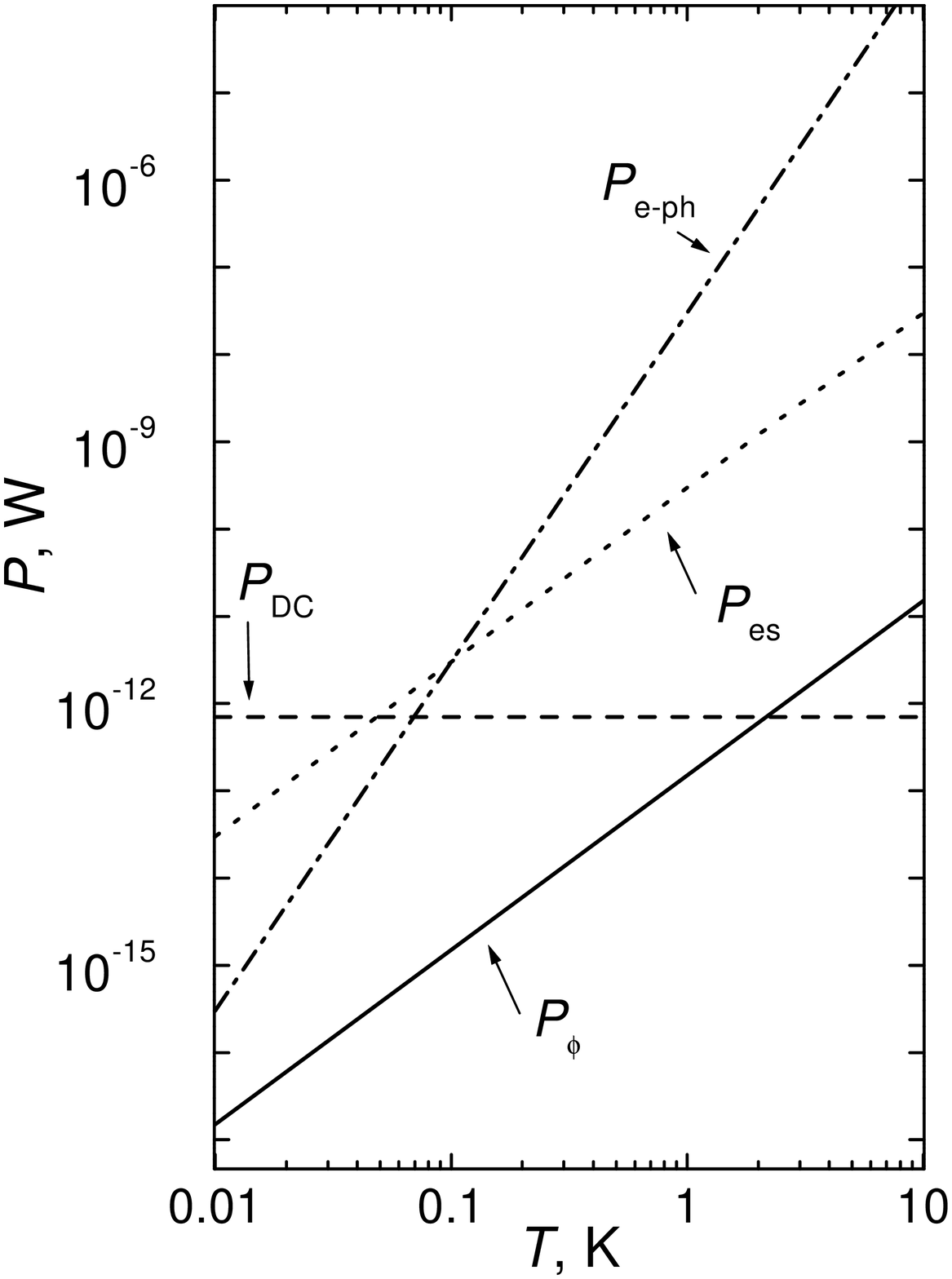, height=0.9\textwidth}
\caption{The temperature dependences of $P_{\varphi}, $ $P_{e-ph},$ and
  $P_{es}$ for sample Au-2~[16]. The horizontal dashed
  line is the power of the DC current that has been used in
  measurements.}
\label{Fig.2}
\end{figure}

We believe that the experimental results reported in
Ref.~\cite{Mohanty} can be explained by the nonequilibrium
external noise. To be specific, let us estimate typical values of
$P_\varphi $, $P_{es}$, and $P_{e-ph}$ for sample Au-2
\cite{Mohanty}: a gold film (thickness $a=60nm$, width $W=110nm$,
length $L=207\mu m$) with the resistance $R=302\Omega $ and
diffusion constant $D=612cm^2/s$. The temperature dependences of
$P_\varphi $ as well as $P_{e-ph}$ and $P_{es}$ calculated for
$\Delta T=0.3T_{ph}$ are shown in Fig.~2; note that $P_\varphi $
is in the sub-picowatt range at $T\leq 0.1K$.  At low
temperatures, electron diffusion is the process controlling
energy flow out of the electron gas, while at higher temperatures
the phonon emission dominates the electron gas cooling. For
conditions of the experiment~\cite{Mohanty}, balancing of the
noise power $P_\varphi $, sufficient for phase breaking, by the
outcoming power due to phonon emission and hot-electron
out diffusion, $P_{e-ph}+P_{es}$, corresponds to a {\it
  negligible} increase of the electron temperature. In this
situation, the rf noise can efficiently {\it destroy} the phase
coherence of the electron wavefunction {\it without} heating the
electron gas. This explains why a well-pronounced temperature
dependence of the resistance (due to the interaction effects) has
been observed at temperatures where the dependence $ L_\varphi
(T)$ was already completely saturated~\cite{Mohanty}.

\begin{table}[h]
\caption{Parameters of the samples studied in
  Ref.~\protect\cite{Mohanty}, ($Au-1$ --- $Au-4$) and the samples
described in Sec.~\protect\ref{sec:5}, ($GaAs$). 
$R $ is the resistance of the
  sample, $L$ is its length, $g=25.8k\Omega /R$ is the
  dimensionless conductance, $\tau _\varphi ^{sat}$ is the
  experimental saturation value of the dephasing time claimed in
  Ref.~\protect\cite{Mohanty}, $V_{DC}$ is the measuring DC bias
  applied to the sample~\protect\cite {thanks}, $P_{DC}$ is the
  corresponding DC power, $V_\varphi $ and $P_\varphi $ is respectively
  the AC voltage and power at optimal frequency $\Omega \simeq
  1/\tau _\varphi ^{sat}$ needed to produce $\tau _\varphi ^{sat}$, see
  Eq.~(\protect\ref{est5}).  $P_\varphi$ and $V_\varphi$ 
for $GaAs$ sample not showing
saturation is the estimate of the $AC$ power required to affect the
observed dependence.}
\label{table1}\center{\
\begin{tabular}{|c|c|c|c|c|c|c|}
\hline
& $Au-1$ & $Au-2$ & $Au-3$ & $Au-4$ & $GaAs$ \\ \hline
$R,\ \Omega$ & $1,687$ & $302$ & $1,443$ & $1,812$ & $9 \times 10^6$ \\
\hline
$L,\ \mu {\rm m}$ & $57.9$ & $207$ & $155$ & $57.9$ & $500$ \\
\hline
$g(L)$ & $15.1$ & $84.1$ & $17.6$ & $14.0$ & $2.9 \times 10^{-3}$
\\ \hline
$g(L_\varphi)$ & $160$ & $1090$ & $525$ & $225$ & 
$ 3 \mbox{---} 10 $
\\ \hline
$\tau_\varphi^{sat}, \ n{\rm s}$ & $3.41$ & $4.19$ & $2.24$ & $1.56$ &
no saturation \\ \hline
$V_{DC}, \ \mu {\rm V}$ & $8.37$ & $14.57$ & $14.68$ & $8.64$  
& $ 50 $ \\ \hline
$P_{DC}, \ {\rm W}$ & $4.2\times 10^{-14}$ & $7.0\times 10^{-13}$ & $
1.5\times 10^{-13}$ & $4.0\times 10^{-14}$ & $2.8
\times 10^{-16}$ \\ \hline
$V_{\varphi}, \ \mu {\rm V}$ & $2.0$ & $2.0$ & $8.8$ & $6.4$ & $2
\times 10^4$
\\
\hline
$P_{\varphi}, \ {\rm W}$ & $2.4\times 10^{-15}$ & $1.4\times 10^{-14}$ & $
5.4\times 10^{-14}$ & $2.2\times 10^{-14}$ & $ > 4 \times 10^{-9} 
$ \\ \hline
\end{tabular}
}
\end{table}

In fact, the dephasing power $P_\varphi$ is even much
smaller than the power dissipated by the DC voltage which was
applied for the measurements~\cite{thanks} (see Fig.~2).
Table~\ref{table1} presents some parameters of the samples
\cite{Mohanty} and the noise power in the frequency range
$f=4\times 10^7\,\mbox{---} \,2\times 10^8{\rm Hz}$, which would
provide the experimentally observed dephasing rate $1/\tau _\varphi
$.  One can see that $P_\varphi $ for most samples is about an order
of magnitude smaller than $P_{DC}$, which is experimentally
proven not to heat the wire.

\section{Conclusion}

In this lecture we argue that there is no need to modify the
existing theory of dephasing in disordered conductors due to the
electron-electron interactions. We have demonstrated that in
$\delta$-doped $GaAs$ structures the theory describes both the
temperature and magnetic field dependences of the conductivity
in the weak localization regime up to the crossover to the strong
localization.

We have proposed a mechanism that may be responsible for
saturation of the dephasing rate at $T \to 0$ in many
experiments. This is dephasing by the external non-equilibrium
noise. It turns out that the dephasing effect of this noise is
much stronger than its heating effect. Therefore, the temperature
dependence of the conductivity observed in Ref.~\cite{Mohanty}
{\em does not prove } the absence of this noise in the system.

In order to preserve the phase coherence for a long time, one has
to reduce the noise amplitude in the frequency range $\Omega \sim
1/\tau_\varphi$ below a very low level. Such reduction could
be a difficult technical problem at ultra-low temperatures.

\end{document}